\documentclass[10pt]{article}
\usepackage{amsmath}
\usepackage{amsthm}
\usepackage{amsfonts}
\usepackage{amssymb}
\usepackage{graphics}
\usepackage{graphicx}
\usepackage{amsbsy}
\newcommand{\LL}{{\mathcal L}}

\newcommand{\R}{{\mathbb R  }}
\newcommand{\RR}{{\mathcal R  }}
\newcommand{\pd}[2]{\frac{\partial #1}{\partial #2}}
\newcommand{\pdt}[1]{\frac{\partial #1}{\partial t}}
\newcommand{\pdtau}[1]{\frac{\partial #1}{\partial \tau}}

\newcommand{\pddd}[3]{\frac{\partial^2 #1}{\partial {#2} \partial{#3}}}

\newcommand{\eps}{\epsilon}

\begin{document}
\setlength{\baselineskip}{10pt}
\title{ A MULTISCALE APPROACH TO BROWNIAN MOTORS}
\author{G.A. Pavliotis \\
        Department of Mathematics\\
    Imperial College London \\
        London SW7 2AZ, UK
                    }
\maketitle

\begin{abstract}
The problem of Brownian motion in a periodic potential, under the influence of
external forcing, which is either random or periodic in time, is studied in this
paper. Multiscale techniques are used to derive general formulae for the steady
state particle current and the effective diffusion tensor. These formulae are then
applied to calculate the effective diffusion coefficient for a Brownian particle in
a periodic potential driven simultaneously by additive Gaussian white and colored
noise. Our theoretical findings are supported by numerical simulations.
\end{abstract}

\section{Introduction}
\label{sec:intro}
Particle transport in spatially periodic, noisy systems has attracted considerable
attention over the last decades, see e.g. \cite[Ch. 11]{Ris84}, \cite{reimann} and
the references therein. There are various physical systems where Brownian motion in
periodic potentials plays a prominent role, such as  Josephson junctions
\cite{BarPater82}, surface diffusion \cite{sancho_al04b, sancho_al04a} and
superionic conductors \cite{FulPieSchnStras75}. While the system of a Brownian
particle in a periodic potential is kept away from equilibrium by an external,
deterministic or random, force, detailed balance does not hold. Consequently, and in
the absence of any spatial symmetry, a net particle current will appear, without any
violation of the second law of thermodynamics.  It was this fundamental observation
\cite{magnasco93} that led to a revival of interest in the problem of particle
transport in periodic potentials with broken spatial symmetry. These types of non--
equilibrium systems, which are often called Brownian motors or ratchets,
 have found new and exciting applications e.g as the basis of theoretical models for
 various intracellular transport processes such as molecular motors \cite{BustKelOst01}.
Furthermore, various experimental methods for particle separation have been
suggested which are based on the theory of Brownian motors \cite{BierAst96}.

The long time behavior of a Brownian particle in a periodic potential is determined
uniquely by the { \it effective drift} and the {\it effective diffusion tensor}
which are defined, respectively, as
\begin{equation}
U_{eff} = \lim_{t \rightarrow \infty} \frac{\langle x(t) - x(0) \rangle}{t}
\label{e:drift}
\end{equation}
and
\begin{equation}
D_{eff} = \lim_{ t \rightarrow \infty} \frac{\langle x(t) - \langle x(t)
 \rangle ) \otimes ( x(t) - \langle x(t)  \rangle ) \rangle}{2 t}.
\label{e:diff}
\end{equation}
Here $x(t)$ denotes the particle position, $\langle \cdot \rangle$ denotes ensemble
average and $\otimes$ stands for the tensor product.  Indeed, an argument based on
the central limit theorem \cite[Ch. 3]{lions}, \cite{kipnis} implies that at
long times the particle performs an effective Brownian motion which is a Gaussian
process, and hence the first two moments are sufficient to determine the process
uniquely. The main goal of all theoretical investigations of noisy, non--equilibrium
particle transport is the calculation of \eqref{e:drift} and \eqref{e:diff}. One
wishes, in particular, to analyze the dependence of these two quantities on the
various parameters of the problem, such as the friction coefficient, the temperature
and the particle mass.

Enormous theoretical effort has been put into the study of Brownian ratchets and,
more generally, of Brownian particles in spatially periodic potentials
\cite{reimann}. The vast majority of all these theoretical investigations is concerned
with the calculation of the effective drift for one dimensional models. This is not
surprising, since the theoretical tools that are currently available are not
sufficient for the analytical treatment of the multi--dimensional problem. This is only
possible when the potential and/or noise are such that the problem can be reduced to a
one dimensional one \cite{EichReim05}. For more general multi--dimensional
problems one has to resort to numerical simulations. There are various applications,
however, where the one dimensional analysis is inadequate. As an example we mention the
technique for separation of macromolecules in microfabricated sieves that was
proposed in \cite{DerAst98}. In the two--dimensional setting considered in this
paper, an appropriately chosen driving force in the $y$ direction produces a
constant drift in the $x$ direction, but with a zero net velocity in the $y$
direction. On the other hand, a force in the $x$ direction produces no drift in the
$y$ direction. The theoretical analysis of this problem requires new technical tools.

Furthermore, the number of theoretical studies related to the calculation of the
effective diffusion tensor has  also been  scarce
\cite{gang_al96,lindner_al01,reimann_al02,reimann_al01,reimann_al98}. In these
papers, relatively simple potentials and/or forcing terms are considered, such as
tilting periodic potentials or simple periodic in time forcing. It is widely
recognized that the calculation of the effective diffusion coefficient is
technically more demanding than that of the effective drift. Indeed, as we will show in
this paper, it requires the solution of a Poisson equation, in addition to the solution
of the stationary Fokker--Planck equation which is sufficient for the calculation of
the effective drift. Diffusive, rather than directed, transport can be potentially
extremely important in the design of experimental setups for particle selection
\cite[Sec 5.11]{reimann} \cite{reimann_al98}. It is therefore desirable to develop
systematic tools for the calculation of the effective diffusion coefficient (or
tensor, in the multi--dimensional setting).

From a mathematical point of view, non--equilibrium systems which are subject to
unbiased noise can be modelled as non--reversible Markov processes \cite{qian02} and
can be expressed in terms of solutions to stochastic differential equations (SDEs).
The SDEs which govern the motion of a Brownian particle in a periodic potential
possess inherent length and time scales: those related to the spatial period of the
potential and the temporal period (or correlation time) of the external driving
force. From this point of view the calculation of the effective drift and the
effective diffusion coefficient amounts to studying the behavior of solutions to the
underlying SDEs at length and time scales which are much longer than the
characteristic scales of the system. A systematic methodology for studying
problems of this type, which is based on scale separation, has been developed many
years ago \cite{lions, papan:asympt_sde_1, papan:asymptot_sde_2}. The techniques
developed in the aforementioned references are appropriate for the asymptotic analysis of
stochastic systems (and Markov processes in particular) which are spatially and/or
temporally periodic. The purpose of this work is to apply these multiscale
techniques to the study Brownian motors in arbitrary dimensions, with particular
emphasis to the calculation of the effective diffusion tensor.

The rest of this paper is organized as follows. In section \ref{sec:model} we
introduce the model that we will study. In section \ref{sec:mult_an} we obtain
formulae for the effective drift and the effective diffusion tensor in the case
where all external forces are Markov processes. In section \ref{sec:ou} we study the
effective diffusion coefficient for a Brownian particle in a periodic potential
driven simultaneously by additive Gaussian white and colored noise. Section
\ref{sec:concl} is reserved for conclusions. In Appendix A we derive
formulae for the effective drift and the effective diffusion coefficient for the
case where the Brownian particle is driven away from equilibrium by periodic in time
external fluctuations. Finally, in appendix B we use the method developed in this paper to calculate 
 the effective diffusion coefficient of an overdamped particle in a one
dimensional tilted periodic potential. 

%
%
%
\section{The Model}
\label{sec:model}

We consider the overdamped $d$--dimensional stochastic dynamics for a state variable
$x(t) \in \R^d$ \cite[sec. 3]{reimann}
\begin{equation}
\gamma \dot{x}(t)= -  \nabla V(x(t),f(t)) + y(t) +\sqrt{2 \gamma k_B T} \xi(t),
\label{e:main}
\end{equation}
where $\gamma$ is the friction coefficient, $k_B$ the Boltzmann constant and $T$
denotes the temperature. $\xi (t)$ stands for the standard $d$--dimensional white noise
process, i.e.
$$
\langle \xi_i(t) \rangle = 0 \quad \mbox{and} \quad \langle \xi_i(t) \xi_j (s) \rangle
= \delta_{ij} \delta(t - s), \quad i, j = 1, \dots d.
$$
We take $f(t)$ and $y(t)$ to be Markov processes with respective state spaces $E_f, \, E_y$
and generators $\LL_f, \, \LL_y$. The potential $V(x,f)$ is periodic in $x$ for every
$f$, with period $L$ in all spatial directions:
$$
V(x + L \hat{e}_i,f) = V(x, f), \quad i=1, \dots, d,
$$
where $\{\hat{e}_i\}_{i=1}^d$ denotes  the standard basis of $\R^d$. We will use the
notation $Q =[0,L]^d$.

The processes $f(t)$ and $y(t)$ can be continuous in time diffusion
processes which are constructed as solutions of stochastic differential equations,
dichotomous noise \cite[Ch. 9]{HorsLef84}, more general Markov chains etc. The (easier) case where
$f(t)$ and $y(t)$ are deterministic, periodic functions of time is treated in the appendix.

For simplicity, we have assumed that the temperature in \eqref{e:main} is constant. However, this
assumption is with no loss of generality, since eqn. \eqref{e:main} with a time dependent
temperature can be mapped to an equation with constant temperature and an
appropriate effective potential \cite[sec. 6]{reimann}. Thus, the above framework is
general enough to encompass most of the models that have been studied in the
literature, such as pulsating, tilting, or temperature ratchets. We remark that the
state variable $x(t)$ does not necessarily denote the position of a Brownian
particle. We will, however,  refer to $x(t)$ as the particle position in
the sequel.

The process $\{ x(t), f(t), y(t) \}$ in the extended phase space $\R^d \times E_f
\times E_y$ is Markovian with generator
$$
\LL = F(x,f,y) \cdot \nabla_x + D \Delta_x + \LL_f + \LL_y,
$$
where $D:= \frac{k_B T}{\gamma}$ and
$$
F(x,f,y) =  \frac{1}{\gamma} \left( -\nabla V(x,f) + y \right).
$$
To this process we can associate the initial value problem for the
backward Kolmogorov Equation \cite[Ch. 8]{oks98}
\begin{equation}
\pdt{u} = \LL u, \quad u(x,y,f,t=0) = u_{in}(x,y,f).
\label{e:bk_mar}
\end{equation}
which is, of course, the adjoint to the Fokker--Planck equation. Our derivation of
formulae for the effective drift and the effective diffusion tensor is based on
singular perturbation analysis of the initial value problem \eqref{e:bk_mar}.
%
%
%
%
\section{Multiscale Analysis}
\label{sec:mult_an}
In this section we derive formulae for the effective drift and the effective
diffusion tensor for $x(t)$, the solution of \eqref{e:main}. Let us outline the basic philosophy
behind the derivation of formulae \eqref{e:drift_ran} and \eqref{e:diff_ran}. We are interested in
the long time, large scale behavior of $x(t)$. For the analysis that follows it is convenient to 
introduce a parameter $\eps \ll 1$ which in effect is the ratio between the length scale defined 
through the period of the potential and a large "macroscopic" length scale at which the motion 
of the particle is governed by and effective Brownian motion. The limit $\eps \rightarrow 0$ 
corresponds to the limit of infinite scale separation. The behavior of the system in this limit 
can be analysed using singular perturbation theory. 

We remark that the calculation of the effective drift and the effective diffusion tensor are
performed seperately, because a different re--scaling is needed in each case. This is due to the
fact that advection and diffusion have different characteristic time scales.

\subsection{Calculation of the Effective Drift}
The backward Kolmogorov equation reads
\begin{equation}
\pdt{u(x,y,f,t)} = \left( F(x,f,y) \cdot \nabla_x + D \Delta_x +
\LL_f + \LL_y  \right) u(x,y,f,t).
\label{e:bk_mar_1}
\end{equation}
We re--scale a space and time in \eqref{e:bk_mar_1} according to
$$
x \rightarrow \eps x, \quad t  \rightarrow \eps t
$$
and divide through by $\eps$ to obtain
\begin{equation}
\frac{\partial u^{\eps}}{\partial t} =  \frac{1}{\eps} \left( F \left((
\frac{x}{\eps},f,y \right) \cdot \nabla_x + \eps D \Delta_x +
 \LL_f +  \LL_y \right) u^{\eps}.
\label{e:bk_ran_1}
\end{equation}
We solve \eqref{e:bk_ran_1} pertubatively by looking for a solution in the form of a
two--scale expansion
\begin{equation}
u^{\eps}(x,f,y,t) = u_0 \left(x,\frac{x}{\eps}, f,y ,t \right) +  \eps
u_1\left(x,\frac{x}{\eps}, f,y ,t \right) + \eps^2 u_2 \left(x,\frac{x}{\eps}, f,y
,t \right) + \dots.
\label{e:exp_ran}
\end{equation}
All terms in the expansion \eqref{e:exp_ran} are periodic functions of $z = x/\eps$. From the
chain rule we have
\begin{equation}
\quad \nabla_x \rightarrow \nabla_x + \frac{1}{\eps} \nabla_z.
\label{e:transf_ran}
\end{equation}
Notice that we do not take the terms in the expansion \eqref{e:transf_ran} to depend explicitly on
$t/\eps$. This is because the coefficients of the backward Kolmogorov equation \eqref{e:bk_ran_1} do 
not depend explicitly on the fast time $t/\eps$. In the case where the fluctuations are periodic,
rather than Markovian, in time, we will need to assume that the terms
in the multiscale expansion for $u^{\eps}(x,t)$ depend explicitly on $t/\eps$. The details are
presented in the appendix. 

We substitute now \eqref{e:exp_ran} into \eqref{e:bk_mar_1}, use
\eqref{e:transf_ran} and treat $x$ and $z$ as
independent variables. Upon equating the coefficients of equal powers in $\eps$ we
obtain the following sequence of equations
\begin{eqnarray}
\LL_0 u_0 & = & 0, \label{e:00_ran} \\
\LL_0 u_1 & = & - \LL_1 u_0 + \pdt{u_0}, \label{e:01_ran} \\
\dots & = & \dots, \nonumber
\end{eqnarray}
where
\begin{equation}
\mathcal{L}_0 = F(z, f, y) \cdot \nabla_z + D \Delta_z + \LL_y +
\LL_f
\label{e:l0}
\end{equation}
and
$$
\LL_1 =  F(z,f,y) \cdot \nabla_x + 2 D  \nabla_z \nabla_x.
$$
The operator $\LL_0$ is the generator of a Markov process on $Q \times E_y \times
E_f$. In order to proceed we need to assume that this process is
ergodic: there exists a unique stationary solution of the Fokker--Planck equation
\begin{equation}
\LL_0^* \rho(z, y,f ) = 0,
\label{e:adj_ran}
\end{equation}
with
$$
\int_{Q \times E_y \times E_f } \rho(z,y, f) \, dz dy d f = 1
$$
and
$$
\LL_0^* \rho = \nabla_z \cdot \left(  F(z, f, y) \rho \right) + D
\Delta_z \rho +  \LL^*_y \rho + \LL^*_f \rho.
$$
In the above $ \LL^*_f$ and $\LL^*_y $ are the Fokker--Planck operators of $f$ and $y$,
respectively. The stationary density $\rho(z,y,f)$ satisfies periodic boundary conditions in
$z$ and appropriate boundary conditions in $f$ and $y$. We emphasize that the ergodicity of
the "fast" process is necessary for the very existence of an effective drift and an
effective diffusion coefficient, and it has been tacitly assumed in all theoretical
investigations concerning Brownian motors \cite{reimann}.

Under the assumption that \eqref{e:adj_ran} has a unique solution eqn. \eqref{e:00_ran}
implies, by Fredholm alternative, that $u_0$ is independent of the fast scales:
$$
u_0 = u(x,t).
$$
Eqn. \eqref{e:01_ran} now becomes
$$
\LL_0 u_1  =  \pdt{u(x,t)} -  F(z,y,f) \cdot \nabla_x u(x,t).
$$
In order for this equation to be well posed it is necessary that the right hand side
averages to $0$ with respect to the invariant distribution $\rho(z,f,y)$. This leads to the
following backward Liouville equation
$$
\pdt{u(x,t)} = U_{eff} \cdot \nabla_x u(x,t),
$$
with the effective drift given by
\begin{eqnarray}
U_{eff} & = & \int_{Q \times E_y \times E_f} F(z,y,f) \rho(z,y, f) \, dz dy d f
\nonumber \\ & = &   \frac{1}{\gamma} \int_{Q \times E_y \times E_f}   \left( -\nabla V(x,f) +
                      y \right) \rho(z,y, f) \, dz dy d f.
\label{e:drift_ran}
\end{eqnarray}
%
%
%
%
\subsection{Calculation of the Effective Diffusion Coefficient}
We assume for the moment that the effective drift vanishes, $U_{eff} = 0$. We perform a 
diffusive  re--scaling in \eqref{e:bk_mar_1}
$$
x \rightarrow \eps x, \quad t \rightarrow \eps^2 t
$$
and divide through by $\eps^2$ to obtain
\begin{equation}
 \frac{\partial u^{\eps}}{\partial t} =   \frac{1}{\eps^2} \left( F \left(\frac{x}{
 \eps},f,y \right)
\cdot \nabla_x + \eps D \Delta_x + \LL_f + \LL_y  \right) u^{\eps},
\label{e:bk_ran_2}
\end{equation}
We go through the same analysis as in the previous subsection to obtain the
following sequence of equations.
\begin{eqnarray}
\LL_0 u_0 & = & 0, \label{e:10ran} \\
\LL_0 u_1 & = & - \LL_1 u_0, \label{e:11ran} \\
\LL_0 u_2 & = & -\LL_1 u_1  - \LL_2 u_0, \label{e:21ran} \\
\dots & = & \dots \nonumber,
\end{eqnarray}
where $ \LL_0$ and $ \LL_1$ were defined in the previous subsection and
$$
\LL_2 = -\pdt{} + D \Delta_x.
$$
Equation \eqref{e:10ran} implies that $u_0 = u(x,t)$. Now \eqref{e:11ran} becomes
$$
\LL_0 u_1  =  - F(z, y, f) \cdot \nabla_x u(x,t).
$$
Since we have assumed that $U_{eff} = 0$, the right hand side of the above equation
belongs to the null space of $\LL_0^*$ and this equation is well posed. Its solution
is
$$
u_1(x,z,f,y,t) = \chi(z,y,f) \cdot \nabla_x u(x,t),
$$
where the auxiliary field $\chi(z,y,f)$ satisfies the Poisson equation
$$
-\LL_0 \chi (z,y,f) = F(z,y,f)
$$
with periodic boundary conditions in $z$ and appropriate boundary conditions in $y$ and $f$.

We proceed now with the analysis of equation \eqref{e:21ran}. The solvability
condition for this equation reads
$$
\int_{Q \times E_y \times E_f}  \left( -\LL_1 u_1  - \LL_2 u_0 \right) \, dz dy d f
= 0,
$$
from which, after some straightforward algebra, we obtain the limiting backward Kolmogorov
equation for $u(x,t)$
$$
\pdt{u(x,t)} = \sum_{i,j=1}^d D^{eff}_{ij} \pddd{u(x,t)}{x_i}{ x_j}.
$$
The effective diffusion tensor is
\begin{eqnarray}
D^{eff}_{ij} & = & D \delta_{ij} +
 \left\langle F_i(z, y, f) \chi^j(z,y,f) \right\rangle_{\rho} + 2 D
\left\langle \pd{\chi^i(z,y, f)}{z_j} \right\rangle_{\rho},
\label{e:diff_ran}
\end{eqnarray}
where the notation $\langle \cdot \rangle_{\rho}$ for the averaging with respect to
the invariant density has been introduced.

The case where the effective drift does not vanish, $U_{eff} \neq 0$, can be reduced to
the situation analyzed in this subsection through a Galilean transformation with
respect to $U_{eff}$ \footnote{In other words, the process $x^(\eps)(t) := \eps \left( x(t/\eps^2) -
\eps^{-2} U_{eff} t \right)$ converges to a mean zero Gaussian process with effective diffusivity 
given by \eqref{e:diff_ran_1}}. The effective diffusion tensor is now given by
\begin{eqnarray}
D^{eff}_{ij} & = & D \delta_{ij}  +
 \left\langle \left( F_i(z, y, f) - U^i_{eff} \right) \chi^j(z,y,f) \right\rangle_{\rho}  \nonumber 
 \\ && + 2 D \left\langle \pd{\chi^i(z,y, f)}{z_j} \right\rangle_{\rho},
\label{e:diff_ran_1}
\end{eqnarray}
and the field $\chi(z,f,y)$ satisfies the Poisson equation
\begin{equation}
- \LL_0 \chi = F(z,y,f) - U_{eff}.
\label{e:cell}
\end{equation}
\section{Effective Diffusion Coefficient for Correlation Ratchets}
\label{sec:ou}
In this section we consider the following model \cite{BartReimHan96, DoerHorsthRiord94}
\begin{subequations}
\begin{equation}
\gamma \dot{x}(t)= -  \nabla V(x(t)) + y(t) +\sqrt{2 \gamma k_B T} \, \xi(t),
\label{e:motion}
\end{equation}
\begin{equation}
 \dot{y}(t)= - \frac{1}{\tau} y(t) +\sqrt{\frac{2 \sigma}{\tau}} \, \zeta(t),
\label{e:ou}
\end{equation}
\label{e:corell}
\end{subequations}
where $\xi(t)$ and $\zeta(t)$ are mutually independent standard $d$--dimensional
white noise processes. The potential $V(x)$ is assumed to be $L$--periodic in all spatial
directions The process $y(t)$ is the $d$--dimensional Onrstein--Uhlenbeck (OU) process
\cite{Gar85} which is a mean zero Gaussian process with correlation function
$$
\langle y_i(t) y_j(s) \rangle = \delta_{ij} \sigma e^{-\frac{|t - s|}{\tau}}, \quad i,j=1,
\dots, d.
$$

Let $z(t)$ denote the restriction of $x(t)$ to $Q = [0, 2 \pi]^d$. The
generator of the Markov process $\{z(t), \, y(t) \}$ is
$$
\LL = \frac{1}{\gamma} (- \nabla_z V(z) + y) \cdot \nabla_z + D \Delta_z 
+ \frac{1}{\tau} \left( -y \cdot \nabla_y + \sigma \Delta_y \right)
$$
with $D:= \frac{k_B T}{\gamma}$. Standard results from the ergodic theory of Markov processes see 
e.g. \cite[ch. 3]{lions} ensure that the process $\{z(t), y(t) \} \in Q \times \R^d$, with 
generator $\LL$ is ergodic and that the unique invariant measure has a smooth density $\rho(y,z)$
with respect to the Lebesgue measure. This is true even at zero temperature
\cite{per_hom_hypoell, MatSt02}. Hence, the results of section \ref{sec:mult_an} apply: the
effective drift and effective diffusion tensor are given by formulae \eqref{e:drift_ran} and
\eqref{e:diff_ran}, respectively. Of course, in order to calculate these quantities
we need to solve equations \eqref{e:adj_ran} and \eqref{e:cell} which take the form:
\begin{eqnarray*}
&&-\frac{1}{\gamma} \nabla_z \cdot \left( (- \nabla_z V(z) + y) \rho(y,z) \right)  +
D \Delta_z \rho(y,z) + \frac{1}{\tau} \big( \nabla_y \cdot (y
\rho(y,z)) \\ &&+ \sigma \Delta_y \rho(y,z) \big) = 0
\end{eqnarray*}
and
\begin{eqnarray*}
&&-\frac{1}{\gamma} (- \nabla_z V(z) + y) \cdot \nabla_z  \chi(y,z)  - D \Delta_z \chi(y,z) \\ && - 
\frac{1}{\tau} \big( - y \cdot \nabla_y \chi(y,z) + \sigma \Delta_y \chi(y,z) \big) = 
\frac{1}{\gamma} (- \nabla_z V(z) + y) - U. 
\end{eqnarray*}
The effective diffusion tensor is positive definite. To prove this, let $e$ be a
unit vector in $\R^d$, define $f = F \cdot e, \, u = U_{eff} \cdot e$ and let $\phi:= e
\cdot \chi$ denote the unique solution of the scalar problem
$$
-\LL \phi = (F - U) \cdot e =:f -u, \quad  \phi(y,z+L) = \phi(y,z), \quad \langle
\phi \rangle_{\rho} = 0.
$$
Let now $h(y,z)$ be a sufficiently smooth function. Elementary computations yield
$$
\LL^* (h \rho) = -\rho \LL h + 2 D \nabla_z \cdot \left(\rho
\nabla_z h \right) +  \frac{2 \sigma}{\tau} \nabla_y \cdot \left( \rho \nabla_y h \right).
$$
We use the above calculation in the formula for the effective diffusion tensor, together
with an integration by parts and the fact that $\langle \phi(y,z) \rangle_{\rho} =
0$, to obtain
\begin{eqnarray*}
e \cdot D_{eff} \cdot e & = & D + \langle f \phi  \rangle_{\rho}
+
D \langle e \cdot \nabla_z \phi \rangle_{\rho} \\
& = &  D  + \langle u \phi  \rangle_{\rho}  - \langle \phi \LL
\phi
\rangle_{\rho} + 2 D \langle e \cdot \nabla_y \phi \rangle_{\rho} \\
& = &  D  +  D  \langle | \nabla_z \phi|^2
\rangle_{\rho} +  2 D   \langle e \cdot \nabla_y \phi
\rangle_{\rho} + \frac{\sigma}{\tau} \langle | \nabla_y \phi|^2  \rangle_{\rho} \\
& = &  D  \langle |e + \nabla_z \phi|^2  \rangle_{\rho} +
\frac{\sigma}{\tau} \langle | \nabla_y \phi|^2  \rangle_{\rho}.
\end{eqnarray*}
From the above formula we see that the effective diffusion tensor is non--negative
definite and that it is well defined even at zero temperature:
$$
e \cdot D_{eff}(T = 0) \cdot e= \frac{\sigma}{\tau} \langle | \nabla_y \phi (T = 0)|^2
\rangle_{\rho}.
$$
Although we cannot solve these equations in closed form, it is possible to calculate
the small $\tau$ expansion of the effective drift and the effective diffusion
coefficient, at least in one dimension. Indeed, a tedious calculation using singular
perturbation theory, e.g. \cite{HorsLef84, per_hom_inert_part} yields
\begin{equation}
U_{eff} = \mathcal{O}(\tau^3),
\label{e:drift_ou}
\end{equation}
and
\begin{equation}
D_{eff} = \frac{L^2}{Z \widehat{Z}} \left( D + \tau \sigma \left( 1 + \frac{1}{\gamma D^2}
\left( \frac{Z_2}{\widehat{Z}} - \frac{Z_1}{Z} \right) \right) \right) +
\mathcal{O}(\tau^2).
\label{e:diff_ou}
\end{equation}
In writing eqn. \eqref{e:diff_ou} we have used the following  notation
\begin{eqnarray*}
&& Z = \int_{0}^L e^{-\frac{V(z)}{D}} \, dz, \quad \widehat{Z} = \int_{0}^L
e^{\frac{V(z)}{D}} \, dz, \\ && Z_1 = \int_{0}^L V(z) e^{-\frac{V(z)}{D}} \, dz,
\quad Z_2 = \int_{0}^L V(z) e^{\frac{V(z)}{D}} \, dz.
\end{eqnarray*}
It is relatively straightforward to obtain the next order correction to \eqref{e:diff_ou};
the resulting formula is, however, too complicated to be of much use.

The small $\tau$ asymptotics for the effective drift were also studied in \cite{
BartReimHan96, DoerHorsthRiord94} for the model considered in this section and in
\cite{DoerDonKlos98, ElstDoer96} when the external fluctuations are given by a
continuous time Markov chain. It was shown in \cite{DoerDonKlos98, ElstDoer96} that,
for the case of dichotomous noise, the small $\tau$ expansion for $U_{eff}$ is valid
only for sufficiently smooth potentials. Indeed, the first non--zero term--of order
$\mathcal{O}(\tau^3)$--involves the second derivative of the potential. Non--smooth
potentials lead to an effective drift which is $\mathcal{O}(\tau^{\frac{5}{2}})$. On
the contrary, eqn. \eqref{e:diff_ou} does not involve any derivatives of the
potential and, hence, is well defined even for non--smooth potentials. On the other
hand, the $\mathcal{O}(\tau^2)$ term involves third order derivatives of the
potential and can be defined only when $V(x) \in C^3(0,L)$.

We also remark that the expansion \eqref{e:diff_ou} is only valid for positive
temperatures. The problem becomes substantially more complicated at zero temperature
because the generator of the Markov process  becomes a degenerate differential
operator at $T = 0$.

Naturally, in the limit
as $\tau \rightarrow 0$ the effective diffusion coefficient converges to its value for
$y \equiv 0$ :
\begin{equation}
D_{eff} = \frac{L^2 D}{Z \widehat{Z}}.
\label{e:deff_v}
\end{equation}
This is the effective diffusion coefficient for a Brownian particle moving in a
periodic potential, in the absence of external fluctuations \cite{lifson_jackson62,
vergassola}. It is well known, and easy to prove, that the effective diffusion
coefficient given by \eqref{e:deff_v} is bounded from above by $D$. This not the
case for the effective diffusivity of the correlation ratchet \eqref{e:corell}.

We compare now the small $\tau$ asymptotics for the effective diffusion coefficient with
Monte Carlo simulations. The results presented in figures
1 and
2 were obtained from the numerical solution of equations \eqref{e:corell}
using the Euler--Marayama method, for the cosine potential $V(x) = \cos(x)$. The integration
step that was used was $\Delta t = 10^{-4}$ and the total number of integration steps was
$10^7$. The effective diffusion coefficient was calculated by ensemble averaging over $2000$
particle trajectories which were initially uniformly distributed on $[0, 2 \pi]$.
\begin{figure}
\begin{center}
\includegraphics[width=2.9in, height = 2.9in]{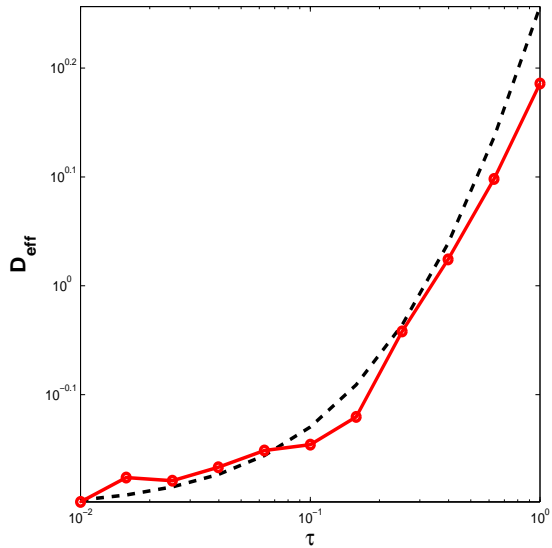}
\caption{Effective diffusivity for \eqref{e:corell} with $V(x) = \cos(x)$ as a function of 
$\tau$, for $\sigma = 1, \, D =  \frac{k_B T}{\gamma} = 1, \, \gamma = 1$. Solid line: Results 
from Monte Carlo simulations. Dashed line: Results from formula \eqref{e:diff_ou}.}
\end{center}
\label{fig:efdif_tau}
\end{figure}
\begin{figure}
\begin{center}
\includegraphics[width=2.9in, height = 2.9in]{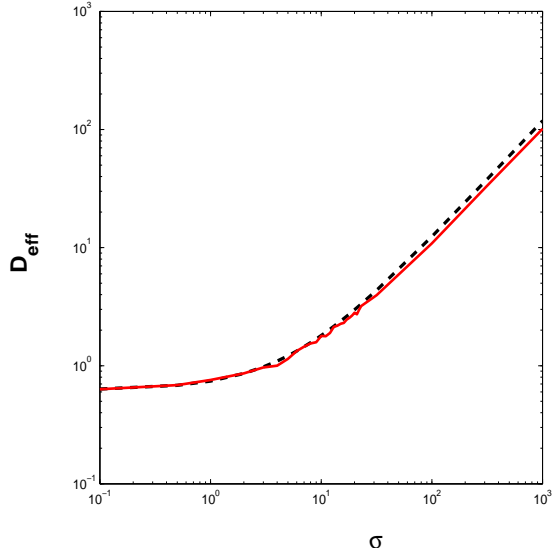}
\caption{Effective diffusivity for \eqref{e:corell} with $V(x) = \cos(x)$  as a function of 
$\sigma$, for $\tau = 0.1, \, D = \frac{k_B T}{\gamma}=1,\, \gamma = 1$. Solid line: Results 
from Monte Carlo simulations. Dashed line: Results from formula \eqref{e:diff_ou}.}
\end{center}
\label{fig:efdif_sigma}
\end{figure}
In figure
1 we present the effective diffusion coefficient as a function of the
correlation time $\tau$ of the OU process. We also plot the results of the small $\tau$
asymptotics. The agreement between theoretical predictions and numerical results is quite
satisfactory, for $\tau \ll 1 $. We also observe that the effective diffusivity is an increasing
function of $\tau$.

In figure 2
we plot the effective diffusivity as a function of the noise
strength  $\sigma$ of the OU process. As expected, the effective diffusivity is an increasing
function of $\sigma$. The agreement between the theoretical predictions from \eqref{e:diff_ou} and
the numerical experiments is excellent.


%
%
%
%
\section{Summary and Conclusions}
\label{sec:concl}
The problem of Brownian motion in a periodic potential and driven by external
forces was studied in this paper. Using multiscale techniques we derived formulae
for the effective drift and the effective diffusion tensor. We used these
formulae to calculate the effective diffusion coefficient for a Brownian particle in
a one dimensional periodic potential which is driven by additive colored noise. The
theoretical predictions are in very good agreement with the results obtained from direct
numerical calculations.

In order to streamline the presentation, we studied the cases of Markovian and
time--periodic external fluctuations separately. It is
straightforward, of course,  to combine the two and obtain formulae for the
effective drift and the effective diffusion coefficient when some of the processes
are random and some periodic. Furthermore, the case where the forcing terms depend
on the particle position can be treated easily within the framework developed in
this paper. In order for our method to be applicable in this case it is necessary for the fast
process to be ergodic for all values of the state variable $x$. We believe, thus, that our
multiscale approach is general enough to encompass essentially all models for periodic
potentials and external fluctuations that have been considered within the framework of
Brownian motors.

The multiscale technique employed in this paper provides us with a systematic method for analyzing 
the long time behavior of solutions to the stochastic equations of motion. The assumption of scale
separation upon which the method is based is always satisfied for systems with a finite number of 
characteristic length and time scales. Naturally, the results of the multiscale analysis are valid at
length and time scales large compared to the largest length and time scales of the system. For
example, the time required for the correlation ratchet \eqref{e:corell} to reach its asymptotic
 Brownian regime increases as the correlation time $\tau$ increases. However, for finite $\tau$ the
 system will always eventually perform an effective Brownian motion.
 
In this paper we only presented formal calculations. We emphasize however that all the results
presented in this paper can be justified rigorously, and in a very general setting. The key
assumptions are the ergodicity of the fast process generated by $\LL_0$, defined in
\eqref{e:l0}, and the well--posedness of the Poisson equation \eqref{e:cell}. Ergodicity implies the
existence of the effective drift: this is a consequence of the strong law of large numbers. 
Ergodicity, together with well--posedness of eqn. \eqref{e:cell} imply the existence of the 
effective diffusion coefficient: this is a consequence of the martingale central limit theorem, see 
e.g. \cite[ch. 7]{ethier_86}. Needless to say, checking the ergodicity of the fast process and the 
well posedness of \eqref{e:cell} are not trivial tasks. Results of this form are proved, for example,
in \cite{per_hom_hypoell}.   

In the case of stochastic external fluctuations, the calculation of the effective diffusion tensor
requires the numerical solution of equations \eqref{e:adj_ran} and \eqref{e:cell}. It is conceivable
that this task might be as computationally
demanding as direct Monte Carlo simulations. From this point of view the main advantage of our
approach is that it provides a natural starting point for careful asymptotic analysis  and
rigorous study of the dependence of the effective drift and the effective diffusion coefficient on
the various parameters of the problem.

On the contrary, when the coefficients of eqn. \eqref{e:main} are periodic in time, the calculation
of $D_{eff}$ requires the numerical solution of eqns. \eqref{e:adj_per} and \eqref{e:cell_per_2}.
These are  PDEs with periodic coefficients and can be
routinely solved with e.g. a spectral method. In this setting, our approach offers a clear
computational advantage over Monte Carlo simulations. We also mention in passing that the
various formulae for the effective diffusion coefficient that have been derived in the literature
 \cite{gang_al96,lifson_jackson62, reimann_al01, reimann_al98} can be obtained from equation
 \eqref{e:diff_per}: they correspond to cases where equations \eqref{e:drift_per} and
 \eqref{e:cell_per_2} can be solved analytically. An example--the calculation of the effective
 diffusion coefficient of an overdamped Brownian particle in a tilted periodic potential--is
 presented in appendix. Similar calculations yield analytical expressions for all other exactly
 solvable models that have been considered in the literature.

In this paper we have studied the overdamped limit. There are various problems, however,
where inertial effects are important and cannot be neglected. As examples we mention
rocking ratchets in SQUIDS, \cite[sec. 5.10]{reimann}, surface diffusion
\cite{sancho_al04b, sancho_al04a}, inertial particles \cite{per_hom_inert_part,
per_hom_hypoell}. Our framework can be generalized to include inertial effects. It is very
interesting, then, to understand the dependence of the effective diffusion coefficient on the
particle inertia. This problem is currently under investigation.

\bigskip

{\bf Acknowledgements} The author is indebted to Professors C.R. Doering and A.M. Stuart for useful
discussions and suggestions.

%
%
%
%


%
%
%
%
\appendix
\section{Time--Periodic Coefficients}
In this appendix we derive formulae for the mean drift and the effective diffusion
coefficient for a Brownian particle which moves according to
\begin{equation}
\gamma \dot{x}(t)= -  \nabla V(x(t),t) + y(t) +\sqrt{2 \gamma k_B T(x(t),t)} \xi(t),
\label{e:append}
\end{equation}
for space--time periodic potential $V(x,t)$ and temperature $T(x,t) > 0$, and periodic in time force
 $y(t)$. We take the spatial period to be $L$ in all directions and the
temporal period of $V(x,t), \, T(x,t)$ and $y(t)$ to be $\mathcal{T}$. We use the notation $Q =
[0,L]^d$. Equation \eqref{e:append}
is interpreted in the It\^{o} sense. This is the correct interpretation when eqn. \eqref{e:append}
is obtained from the full phase space dynamics, in the limit as the inertia tends to $0$
\cite{paper3_stuart, paper2_stuart}. Since the analysis is very similar to the one presented in
\cite{lions, vergassola} we will be brief.
\subsection{Calculation of the Effective Drift}
The backward Kolmogorov equation corresponding to \eqref{e:append} reads
\begin{equation}
\pdt{u(x, t)} = F(x,t) \cdot \nabla_x u(x,t) + \frac{k_B T(x, t)} {\gamma} \Delta_x
u(x,t),
\label{e:bk_per}
\end{equation}
where $F(x,t) = \frac{1}{\gamma} (- \nabla V(x,y) +y(t))$. We re--scale space and time in
\eqref{e:bk_per} according to $x = \rightarrow \eps x, \, t \rightarrow \eps t$. We look for a
solution of the resulting equation of the form
\begin{equation}
u^{\eps}(x,t) = u_0 \left(x,\frac{x}{\eps}, t, \frac{t}{ \eps} \right) +  \eps
u_1\left(x,\frac{x}{\eps}, t, \frac{t}{ \eps} \right) + \eps^2
u_2\left(x,\frac{x}{\eps}, t, \frac{t}{ \eps} \right) + \dots.
\label{e:exp_per}
\end{equation}
We treat $x, \, z=x/\eps$ and $t, \, \tau = t/\eps$ as independent variables. We
emphasize that $u_i(x,z,t,\tau), \, i =0,1, 2, \dots $ are periodic functions in
{\it both} $z$ and $\tau$. Upon substituting \eqref{e:exp_per} into \eqref{e:bk_per},
using the chain rule and equating the coefficients of equal powers in $\eps$ we
obtain the following sequence of equations
\begin{eqnarray}
\RR_0 u_0 & = & 0, \label{e:00} \\
\RR_0 u_1 & = & - \RR_1 u_0, \label{e:01} \\
\dots & = & \dots, \nonumber
\end{eqnarray}
where
$$
\mathcal{R}_0 = \pdtau{} -\LL_0, \quad  \LL_0 = F(z, \tau) \cdot \nabla_z + \frac{k
T(z,\tau)}{\gamma} \Delta_z
$$
and
$$
\mathcal{R}_1 = \pdt{} -\LL_1, \quad \LL_1 =  F(z, \tau) \cdot
\nabla_x + \frac{2 k_B T(z, \tau)}{\gamma} \nabla_z \nabla_x.
$$
The equation
\begin{equation}
\RR_0^* \rho(z, \tau) = \pdtau{\rho(z, \tau)} + \nabla_z \left( F(z, \tau) \rho(z,
\tau) \right) + \frac{k_B T(z, \tau)}{\gamma} \Delta_z \rho(z, \tau) = 0,
\label{e:adj_per}
\end{equation}
with
$$
\int_0^{\mathcal{T}} \int_{Q} \rho(z, \tau) \, dz d \tau = 1
$$
and periodic boundary conditions in both $z$ and $\tau$ has a unique solution under
appropriate regularity assumptions on $F(z,t), \, T(z,t)$ \cite[Thm. 3.10.1]{lions}.
By Fredholm alternative, \eqref{e:00} implies that
$$
u_0 = u(x,t).
$$
Eqn. \eqref{e:01} now becomes
$$
\RR_0 u_1  =  \pdt{u(x,t)} - F(z, \tau) \cdot \nabla_x u(x,t).
$$
The solvability condition for this equation leads to the following backward
Liouville equation
$$
\pdt{u(x,t)} = U_{eff} \cdot \nabla_x u(x,t),
$$
with
\begin{equation}
U_{eff} = \int_0^{\mathcal{T}} \int_{Q} F(z, \tau) \rho(z, \tau) \, dz d \tau = 0.
\label{e:drift_per}
\end{equation}
%
%
\subsection{Calculation of the Effective Diffusion Coefficient}
Without loss of generality we assume that the effective drift given by \eqref{e:drift_per}
vanishes, $U_{eff} =0$. We perform the diffusive re--scaling in equation \eqref{e:bk_per},
$x \rightarrow \eps x, \, t \rightarrow \eps^2 t$. We go through the same analysis as in the
previous subsection to obtain the following sequence of equations.
\begin{eqnarray}
\RR_0 u_0 & = & 0, \label{e:10} \\
\RR_0 u_1 & = & - \LL_1 u_0, \label{e:11} \\
\RR_0 u_2 & = & -\LL_1 u_1  - \LL_2 u_0 \label{e:21} \\
\dots & = & \dots, \nonumber
\end{eqnarray}
where $\RR_0, \, \LL_0, \, \LL_1$ were defined in the previous subsection and
$$
\LL_2 = \pdt{} - \frac{k_B T(z, \tau)}{\gamma} \Delta_x.
$$
Equation \eqref{e:10} gives that $u_0 = u(x,t)$. Equation \eqref{e:11} becomes
$$
\RR_0 u_1  =  - F(z, \tau) \cdot \nabla_x u(x,t).
$$
Since we have assumed that $U_{eff} = 0$, this equation is well posed. Its solution is
$$
u_1(x,z,t,\tau) = \chi(z, \tau) \cdot \nabla_x u(x,t),
$$
where
\begin{equation}
-\RR_0 \chi (z,\tau) = F(z, \tau).
\label{e:cell_per_1}
\end{equation}
The solvability condition for equation \eqref{e:21} reads
$$
\int_0^{\mathcal{T}} \int_{Q}  \left( -\LL_1 u_1  - \LL_2 u_0 \right) \, dz d \tau =
0,
$$
from which, after some algebra we obtain an evolution equation for $u(x,t)$
$$
\pdt{u(x,t)} = \sum_{i,j=1}^d D_{ij} \pddd{u(x,t)}{x_i} {x_j}.
$$
The effective diffusion coefficient is
\begin{eqnarray}
D_{ij} & = & \delta_{ij} \frac{k_B}{\gamma} \langle T(z, \tau) \rangle_{\rho}
+ \langle F_i(z, \tau) \chi^j(z, \tau) \rangle_{\rho}
\nonumber \\  &&
 + \frac{2 k_B}{\gamma}
\left\langle T(z,\tau) \pd{\chi^i(z, \tau)}{y_j}  \right\rangle_{\rho}
\label{e:diff_per}
\end{eqnarray}
with
$$
\langle \cdot \rangle_{\rho} = \int_0^{\mathcal{T}} \int_{Q} \cdot \rho(z, \tau) \, dz d \tau.
$$
When $U_{eff} \neq 0$, eqns. \eqref{e:cell_per_1} and \eqref{e:diff_per} become, respectively,
\begin{equation}
-\RR_0 \chi (z,\tau) = F(z, \tau) - U_{eff}
\label{e:cell_per_2}
\end{equation}
and
\begin{eqnarray}
D_{ij} & = & \delta_{ij} \frac{k_B}{\gamma} \langle T(z, \tau) \rangle_{\rho}
+ \langle \left( F_i(z, \tau) - U_{eff}^i \right) \chi^j(z, \tau) \rangle_{\rho}
\nonumber \\  &&
 + \frac{2 k_B}{\gamma}
\left\langle T(z,\tau) \pd{\chi^i(z, \tau)}{y_j}  \right\rangle_{\rho}
\label{e:diff_per_1}
\end{eqnarray}
%
%
%
%
%
%
\section{Effective Diffusion Coefficient for Tilted Periodic Potentials}
In this appendix we use our method to obtain a formula for the effective diffusion coefficient of an
overdamped particle moving in a one dimensional tilted periodic potential. This formula was
first derived and analyzed in \cite{reimann_al01, reimann_al02} without any appeal to multiscale
analysis. The equation of motion is
\begin{equation}
\dot{x} = -V'(x) + F + \sqrt{2 D} \xi,
\label{e:tilted}
\end{equation}
where $V(x)$ is a smooth periodic function with period $L$, $F$ and $D >0$ constants and $\xi(t)$ 
standard white noise in one dimension. To simplify the notation we have set $\gamma = 1$. 

The stationary Fokker--Planck equation corresponding to\eqref{e:tilted} is
\begin{equation}
\partial_x \left( \left(  V'(x) - F \right) \rho(x) + D \partial_x \rho(x) \right) = 0,
\label{e:fp_tilted}
\end{equation}
with periodic boundary conditions. Formula \eqref{e:drift_ran} for the effective drift now becomes
\begin{equation}
U_{eff} = \int_0^L (-V'(x) + F) \rho(x) \, dx.
\label{e:drift_tilted}
\end{equation}
The solution of eqn. \eqref{e:fp_tilted} is \cite[Ch. 9]{Straton67}
\begin{equation}
\rho(x) = \frac{1}{Z} \int_x^{x+L} dy Z_+(y) Z_-(x),
\label{e:dens_tilted}
\end{equation}
with
$$
Z_{\pm}(x) := e^{\pm \frac{1}{D} (V(x) - F x)},
$$
and
\begin{equation}
Z = \int_0^L dx \int_x^{x+L} dy Z_+(y) Z_-(x).
\label{e:partition_tilted}
\end{equation}

Upon using \eqref{e:dens_tilted} in \eqref{e:drift_tilted} we obtain \cite[Ch. 9]{Straton67}
\begin{equation}
U_{eff} = \frac{D L}{Z} \left(1  - e^{-\frac{F \, L}{D}} \right).
\label{e:drift_tilted_2}
\end{equation}
Our goal now is to calculate the effective diffusion coefficient. For this we first need to solve the
Poisson equation \eqref{e:cell} which now becomes
\begin{equation}
\LL \chi(x) : = D \partial_{x x} \chi(x) + (-V'(x) + F) \partial_x \chi = 
V'(x) - F + U_{eff},
\label{e:cell_tilted}
\end{equation}
with periodic boundary conditions. Then we need to evaluate the integrals in \eqref{e:diff_ran}:
$$
D_{eff} = D + \int_0^L (- V'(x) + F - U_{eff}) \rho(x) \, dx + 2 D \int_0^L \partial_x 
\chi(x) \rho(x) \, dx. 
$$
It will be more convenient for the subsequent calculation to rewrite the above formula for the 
effective diffusion coefficient in a different form. The fact that $\rho(x)$ solves the stationary
Fokker--Planck equation, together with elementary integrations by parts yield that, for all 
sufficiently smooth periodic functions $\phi(x)$,
$$
\int_0^L \phi(x) (-\LL \phi(x)) \rho(x) \, dx = D \int_0^L (\partial_x \phi(x)  )^2\rho(x) 
\, dx.
$$
Now we have
\begin{eqnarray}
D_{eff} & = & D + \int_0^L (- V'(x) + F - U_{eff}) \chi(x) \rho(x) \, dx + 2 D \int_0^L \partial_x 
              \chi(x) \rho(x) \, dx  
	      \nonumber \\ & = &
              D + \int_0^L ( -\LL \chi(x) ) \chi(x) \rho(x) \, dx + 2 D \int_0^L \partial_x 
              \chi(x) \rho(x) \, dx 
	      \nonumber \\ & = &
              D + D \int_0^L \left( \partial_x \chi(x) \right)^2 \rho(x) \, dx + 2 D \int_0^L 
	      \partial_x  \chi(x) \rho(x) \, dx 	 
	      \nonumber \\ & = &
              D \int_0^L \left(1 + \partial_x \chi(x) \right)^2 \rho(x) \, dx .
\label{e:diff_tilted}  	 	           
\end{eqnarray}
Now we solve the Poisson equation \eqref{e:cell_tilted} with periodic boundary conditions. We
multiply the equation by $Z_-(x)$ and divide through by $D$ to rewrite it in the form
$$
\partial_x (\partial_x \chi(x) Z_-(x)) = - \partial_x Z_-(x) + \frac{U_{eff}}{D} Z_-(x).
$$
We integrate this equation from $x-L$ to $x$ and use the periodicity of $\chi(x)$ and $V(x)$
together with formula \eqref{e:drift_tilted_2} to obtain
$$
\partial_x \chi(x) Z_-(x) \left(1  - e^{-\frac{F \, L}{D}} \right) = - Z_-(x) \left(1  - 
e^{-\frac{F \, L}{D}} \right) + \frac{L}{Z} \left(1  - e^{-\frac{F \, L}{D}} \right) \int_{x-L}^x
Z_-(y) \, dy,
$$
from which we immediately get
$$
\partial_x \chi(x) +1 = \frac{1}{Z} \int_{x-L}^x Z_-(y) Z_+(x) \, dy.
$$
Substituting this into \eqref{e:diff_tilted} and using the formula for the invariant distribution
\eqref{e:dens_tilted} we finally obtain
\begin{equation}
D_{eff} = \frac{D}{Z^3} \int_0^L (I_+(x))^2 I_-(x) \, dx, 
\label{e:diff_tilted_2}
\end{equation}
with
$$
I_+(x) = \int_{x-L}^x Z_-(y) Z_+(x) \, dy \quad \mbox{and} \quad I_-(x) = \int_{x}^{x+L} Z_+(y) 
Z_-(x) \, dy.
$$
Formula \eqref{e:diff_tilted_2} for the effective diffusion coefficient (formula (22) in 
\cite{reimann_al02}) is the main result of this appendix. 

\def\cprime{$'$} \def\cprime{$'$} \def\cprime{$'$}
  \def\Rom#1{\uppercase\expandafter{\romannumeral #1}}\def\u#1{{\accent"15
  #1}}\def\Rom#1{\uppercase\expandafter{\romannumeral #1}}\def\u#1{{\accent"15
  #1}}

\end{document}